\documentclass[12pt]{article}

\usepackage{amsmath}
\usepackage{amssymb}
\usepackage{latexsym}
\usepackage{graphics}
\usepackage{psfrag,fancyhdr,epsfig}

\begin{document}

\newcommand{\newc}{\newcommand}
\newc{\be}{\begin{equation}}
\newc{\ee}{\end{equation}}
\newc{\ba}{\begin{eqnarray}}
\newc{\ea}{\end{eqnarray}}
\newc{\bea}{\begin{eqnarray*}}
\newc{\eea}{\end{eqnarray*}}
\newc{\D}{\partial}
\newc{\ie}{{\it i.e.} }
\newc{\eg}{{\it e.g.} }
\newc{\etc}{{\it etc.} }
\newc{\etal}{{\it et al.}}
\newcommand{\nn}{\nonumber}
\newc{\ra}{\rightarrow}
\newc{\lra}{\leftrightarrow}
\newc{\lsim}{\buildrel{<}\over{\sim}}
\newc{\gsim}{\buildrel{>}\over{\sim}}
\def \omm  {\Omega_{0 {\rm m}}}

\title{Cosmic Acceleration Data and Bulk-Brane Energy Exchange}
\author{C. Bogdanos, S. Nesseris, L. Perivolaropoulos and K. Tamvakis \\
Physics Department, University of Ioannina, Ioannina GR451 10,
Greece}
\maketitle
\begin{abstract}
We consider a braneworld model with bulk-brane energy exchange.
This allows for crossing of the $w=-1$ phantom divide line without
introducing phantom energy with quantum instabilities. We use the
latest SnIa data included in the Gold06 dataset to provide an
estimate of the preferred parameter values of this braneworld
model. We use three fitting approaches which provide best fit
parameter values and hint towards a bulk energy component that
behaves like relativistic matter which is propagating in the bulk
and is moving at a speed $v$ along the fifth dimension, while the
bulk-brane energy exchange component corresponds to negative
pressure and signifies energy flowing from the bulk into the
brane. We find that the best fit effective equation of state
parameter $w_{eff}$ marginally crosses the phantom divide line
$w=-1$. Thus, we have demonstrated both the ability of this class
of braneworld models to provide crossing of the phantom divide and
also that cosmological data hint towards {\textit{natural}} values
for the model parameters.

\end{abstract}

\section{Introduction}
Recent observations~\cite{Riess:2006fw,R} of distant type Ia
supernova (SnIa) provide strong indication of a rather
unconventional cosmic expansion profile for our universe at late
times. It is suggested that we are currently in an accelerated
expansion phase \cite{Riess:2006fw,Astier:2005qq}, contrary to
what we would expect in a matter dominated universe, where gravity
tends to slow down the expansion.   This result has had dramatic
consequences on cosmological models, leading to the assumption of
the existence of {\it dark energy}, an unknown component of the
cosmic energy content which is responsible for the observed
acceleration. Many models have been proposed to explain the nature
of dark energy. These include new forms of matter and energy, like
cosmological constant, scalar fields (quintessence), phantom
fields, hessence, k-essence etc., as well as theories of modified
gravity, extra dimensions and brane models \cite{A,AHDD,DGP,RS}
(see \cite{PHANT,QUINT,KMP,SS,shtanov} and \cite{COSMO} for a list
of possible dark energy models). Apart from confirming the
presence of dark energy, SnIa data also provide valuable
information about its characteristics, namely the equation of
state it obeys. Detailed analysis of various available datasets
implies that a redshift dependent equation of state parameter
$w(z)$ crossing the phantom divide line $w=-1$ is consistent with
the data \cite{NP1} and may be favored by some datasets. If such
crossing is verified, it will render a number of cosmological
models as non-viable candidates for dark energy, as it happens
with pure scalar field models of phantom or quintessence fields.
Obtaining models which can account for both the late times
accelerated expansion, as well as an equation of state parameter
with $w=-1$ crossing becomes thus a high priority.

In \cite{BT} a model was presented based on brane cosmology, where
dark energy is a manifestation of extra dimensions and the
existence of bulk matter and energy exchange between the brane and
the bulk. The model yields accelerated expansion at late times, as
well as an effective equation of state parameter $w_{eff}$, which
crosses the $w_{eff}=-1$ line. Modelling the bulk content as a
slowly moving perfect fluid, we showed that the crossing can occur
without invoking matter that violates the weak energy condition,
provided that the dark radiation field enters with a negative sign
contribution. This requirement can be bypassed if a more general
ansatz is adopted. A similar but qualitatively different from a
physical viewpoint situation occurs in models which consider a
dark matter - dark energy interaction (see \cite{dmdeiter} for
some examples). In \cite{BDT} we departed from the bulk fluid
interpretation and assumed that the bulk pressure and energy
exchange contributions are the dominant ones, while the dark
radiation field is negligible in the range $0 \le z \le 1$. We
thus derived constraints on the allowed range of parameters for
the bulk pressure and energy exchange terms by demanding the model
to conform to the general expected temporal behavior
\cite{SS,Feng}. Our goal in this paper is to perform a more
detailed analysis of the allowed range of parameters by fitting
the model directly to the Gold supernova Ia dataset and checking
whether the best fit values indicate accelerated expansion and
$w=-1$ crossing, instead of requiring such a behavior a priori. In
doing so, we will also keep the dark radiation term and see
whether there are solutions which yield a positive sign for it.

We should note here that the back-reaction of cosmological
perturbations on large scales can mimic the effects of background
accelerated expansion \cite{Kolb:2005da}, thus altering the
results of any analysis that focuses only on the background
evolution. We have chosen not to include the effects of
perturbations in our analysis for two reasons:

\begin{itemize}
    \item The set of equations of cosmological perturbations
are known to be non-closed on the brane \cite{Maartens:2000fg},
i.e. information about the complete bulk evolution and dynamics is
necessary in order to obtain a solution. Since the model we are
considering doesn't originate from an exact bulk solution, but
rather from a reasonable ansatz for the bulk matter content, a
perturbative treatment cannot be particularly helpful.
    \item It turns out that at late times, when the Hubble radius is much
greater than the bulk radius of curvature, the results of
perturbations appear to be equivalent to those obtained when
treating a conventional 4D cosmology \cite{Koyama:2000cc}.
\end{itemize}

\section{General Framework}
We will briefly review the model presented in \cite{BT,BDT}
and we will recast its equations in a form which is more
convenient for the SNIa fit procedure. The model consists of a
four-dimensional brane imbedded inside a five-dimensional bulk
space. Both the bulk and the brane are allowed to carry some
energy content, while there may also be an energy exchange term
between the two. Einstein's equations in this setup are
\begin{equation}
{\cal{G}}_{MN}\equiv\,R_{MN}  - \frac{1}
{2}G_{MN} R \,=\,\frac{1}{4M^3}\,T_{MN}\, {\label{Einstein1}}.
\end{equation}
where $M$ is the $5D$ Planck mass. The energy-momentum tensor
$T_{MN}$ is
\begin{equation}
T_{MN}\,=\,T^{( B )} _{MN} \, +\, T^{( b )} _{MN} \, - G_{MN} \Lambda
\,-g_{\mu\nu}\, \sigma \,\delta ( y
)\,\delta_M^{\mu}\delta_N^{\nu}\,{\label{e-m1}} ,
\end{equation}
The first and second term are the bulk and brane content
respectively, while the third comes from the bulk cosmological
constant and the fourth is the brane tension term. The metric
ansatz we use to study the cosmology of a
Friedmann-Robertson-Walker brane is
\begin{equation}
G_{MN}\,=\,\left(\begin{array}{ccc}
-n^2(y,t)\,&\,0\,&\,0\\
0\,&\,a^2(y,t)\gamma_{ij}\,&\,0\\
0\,&\,0\,&\,b^2(y,t)
\end{array}\right)\,,{\label{metric}}
\end{equation}
where $\gamma_{ij}$ is the metric for the maximally symmetric
3-space on the brane. For a {\textit{bulk energy-momentum tensor}}
$T_{MN}^{(B)}$ we shall take the general matrix
\begin{equation}
{T^{(B)}}^{M}_{\,\,N}\,=\,\left(\begin{array}{ccc}
-\rho_B\,&\,0\,&\,P_5\\
\,0\,&\,P_B\delta^i_{\,\,j}\,&\,0\\
-\frac{n^2}{b^2}P_5\,&\,0\,&\,\overline{P}_B
\end{array}\right)\,\,,\,\,\,
T_{MN}^{(B)}\,=\,\left(\begin{array}{ccc}
\rho_Bn^2\,&\,0\,&\,-n^2P_5\\
\,0\,&\,P_Ba^2\gamma_{ij}\,&\,0\\
-n^2P_5\,&\,0\,&\,\overline{P}_Bb^2
\end{array}\right)\,{\label{bulk-e-m}} .
\end{equation}
Note the presence of the off-diagonal component $T^0_{\,\,5}=P_5$
signifying the flow of energy towards (or from) the brane. Note
also the anisotropic choice $\overline{P}_B\neq P_B$, in general.
Similarly, for a {\textit{brane energy-momentum tensor}}
$T_{MN}^{(b)}$ we shall
 take the matrix
\begin{equation}
{T^{(b)}}^{M}_{\,\,N}\,=\,\frac{\delta(y)}{b}\left(\begin{array}{ccc}
-\rho\,&\,0\,&\,0\\
\,0\,&\,p\delta^i_{\,\,j}\,&\,0\\
\,0\,&\,0\,&\,0
\end{array}\right)\,\,,\,\,\,
T_{MN}^{(b)}\,=\,\frac{\delta(y)}{b}\left(\begin{array}{ccc}
\rho n^2\,&\,0\,&\,0\\
\,0\,&\,pa^2\gamma_{ij}\,&\,0\\
\,0\,&\,0\,&\,0
\end{array}\right)\,.\,{\label{brane-e-m}}
\end{equation}
Following \cite{BT,BDL,KKTTZ,CGW}, we can use the junction
conditions at the brane to derive the effective Friedmann equation
on the brane that determines the cosmological evolution in four
dimensions. Then, using an appropriate ansatz for the bulk
pressure and the energy exchange term and assuming that the brane
energy density is low, so we can neglect quadratic terms in
$\rho$, we can exactly solve the system of equations to find the
Hubble parameter. The ansatz we impose for the bulk
energy-momentum tensor components which enter the cosmological
equations on the brane is
\begin{equation}
\overline{P}_B\,=\,D\,a^{\nu}\,\,,\,\,\,\,P_5\,=\,F\,\left(\frac{\dot{a}}{a}\right)\,a^{\mu}\,.
{\label{Ansatz}}
\end{equation}
and it can be physically justified once we model the bulk content as
a relativistic fluid, slowly moving along the fifth dimension. The
case $\mu=\nu$ corresponds to a slowly moving fluid with equation of
state parameter $\omega=-1-\mu/3$. Note that more than one "fluid"
component may be present. In such a case,
$\overline{P}_B\,=\,\sum_jD_j\,a^{\nu_j}$ and
$P_5\,=\,\frac{\dot{a}}{a}\,\sum_jF_j\,a^{\mu_j}$. Then, it is
possible that $D_1\gg D_i$, while $F_1 \ll F_i$ ($i\neq 1$)and therefore,
at late times $\overline{P}_B\,\approx\,D_1\,a^{\nu_1}$ and
$P_5\,\approx\,F_2\,\left(\frac{\dot{a}}{a}\right)a^{\mu_2}$,
accounting for different exponents.

Substituting these expressions, we arrive at the effective
Friedmann equation
\begin{equation}
\left(\frac{\dot{a}}{a}\right)^2\,+\,\frac{k}{a^2}\,=\,\frac{8\pi}{3}G_N\,\rho_{eff}\,,
{\label{Fried}}
\end{equation}
where $G_N=3\gamma/4\pi=3\sigma/4\pi(24M^3)^2$ is the $4D$ Newton's constant and the
{\textit{effective energy density}} $\rho_{eff}$ stands for
\begin{equation}
\rho_{eff}\,=\,\frac{\tilde{\cal{C}}}{a^{3(1+w)}}\,+\,\frac{{\cal{C}}/2\gamma}{a^4}\,
-\frac{2\delta}{\gamma(\nu+4)}a^{\nu}\,
+\,\frac{2(3w-1)F}{(\mu+4)\left[3(1+w)+\mu\right]}a^{\mu}\,.{\label{rhoeff}}
\end{equation}
where $\gamma \equiv \sigma (24M^3)^{-2}$ and we have defined
$\delta\equiv D/24M^3$. We have assumed the equation of state
$p=w\rho$ for matter on the brane. The constant $\tilde{\cal{C}}$
is just an integration constant for the energy density on the
brane $\rho(a)$. The integration constant ${\cal{C}}$ multiplies
the {\textit{dark radiation} term and is related to global
properties of the the bulk. The third and the fourth term in the
expression for $\rho_{eff}$ come from the bulk pressure $\bar P_B$
and the brane-bulk energy exchange term $P_5$ respectively. The
values of the parameters $\mathcal{C}$, $D$, $F$, $\nu$ and $\mu$,
related to the dark radiation term and the bulk contributions are
crucial in determining the profile of cosmic expansion.

In \cite{BDT} we performed an analysis for the allowed range of
values for the latter four parameters, requiring a cosmological
evolution that conforms to the general profile suggested by
observational data, based on the assumption that the dark
radiation term is negligible at late times. Here we shall perform
a more general analysis, keeping also the dark radiation
contribution and performing a fit to the Gold dataset, thus
obtaining best fit values for the five parameters. In order to
simplify the procedure, we will rewrite the effective Friedmann
equation in terms of energy density parameters, assuming also a
flat space ($k=0$). It should be noted that actually there is an
interesting degeneracy between spatial curvature and time-varying
dark energy as pointed out in Refs
\cite{Nakamura:1998mt,Clarkson:2007bc}. Therefore a small spatial
curvature term could wipe out the mild trend for w=-1 crossing
indicated by the Gold06 data in the context of a flat Universe.
However, we have chosen not to allow for a possibility of non-zero
spatial curvature for two reasons.

\begin{itemize}
    \item There is significant theoretical motivation coming from
inflation that k=0 to a very high accuracy. The possibility of
non-zero spatial curvature would practically invalidate the
flatness problem resolution provided by inflation.
    \item Allowing for of non-zero spatial curvature would introduce one
more parameter to be determined by the SnIa data. However, the
number of parameters involved in our brane world model is already
large and this implies a large error region (see Fig. 1). The
introduction of one more parameter would further extend this
region making the results of our analysis very hard to interpret.
\end{itemize}

Doing so we obtain the following expression
\begin{equation}
\frac{H^2}{H_0^2}\,=\, \Omega _{b} \,a^{-3(1+w)}\,+\,
\Omega_{DR}\,a^{ - 4} \,+\,
  \Omega _B\,a^{\nu} \,  +\,
   \Omega_5\,a^{\mu}\,, \label{H2}
\end{equation}
where we have defined the energy density parameters for brane matter, dark radiation
 and bulk components as
\begin{equation}
\Omega _m \left( {\tilde C} \right) \equiv \frac{{8\pi G_N }}{{3H_0 ^2 }}\tilde C\,,\,\,\,\,\,\,\,
\Omega _{DR} \left( C \right) \equiv \frac{C}{{H_0 ^2 }}\,\label{OmegaB}\,,
\end{equation}
\begin{equation}
\Omega _B \left( {D,\nu } \right) \equiv  - \frac{{4D}}{{24M^3 H_0 ^2 \left( {\nu  + 4} \right)}}\,\,,
\end{equation}
and \begin{equation}
\Omega _5 \left( {F,\mu } \right) \equiv \frac{{8\pi G_N }}{{3H_0 ^2 }}\frac{{2\left( {3w - 1} \right)F}}{{\left( {3\left( {1 + w} \right) + \mu } \right)\left( {\mu  + 4} \right)}}\,.
\end{equation}

The brane energy density parameter $\Omega_m$ corresponds to the
observed matter density. It has a value of approximately $\Omega
_{m}  \simeq 0.27$ \cite{CMB}. Since the brane content is matter
dominated, we will assume a value of $w=0$ in the above
expressions. The four density parameters are in fact related by
the flatness requirement, which yields
\begin{equation}
\Omega_{5} \, = \,1 \,- \Omega_{m} \, -\Omega _{DR}\,
-\Omega_B\,.
\end{equation}
Thus, we can eliminate the $ \Omega _{5}$ term in favor of $\Omega
_{B}$. We see that it depends on both the coefficient $D$, which
characterizes the bulk pressure, as well as the power $\nu$. In
order to make the dependence on $\nu$ explicit, we could rewrite
it as  $\Omega_B\,=\,\frac{\tilde{\Omega}_B}{\nu+4}$.

Using the above parametrization, we are now in position to fit the
model to the supernova Ia data. There are four parameters to fit,
$\Omega _{DR} $,  $\tilde{\Omega}_B$, $\nu$ and $\mu$. Having
determined the best fit values for these parameter, we can infer
the temporal behavior of the deceleration parameter $q$, as well
as the profile of the effective equation of state parameter
$w_{eff}$. We refer to it as an effective parameter, since it
isn't directly related to some particular form of matter, but is
representative of the accumulated effects of both bulk matter and
energy exchange, together with dark radiation due to the presence
of extra dimensions. The parameter $w_{eff}$ is according to the
prescription of \cite{LJ}
\begin{equation}
w_{eff}^{(D)}\,=\,-1\,-\frac{1}{3}\frac{d\ln (\delta H^2)}{d\ln a}\,,{\label{eosp}}
\end{equation}
where $\delta H^2=H^2/H_0^2-\Omega_ma^{-3}$ accounts for all terms in the Friedmann
equation not related to the brane matter. Substituting the above parametrization,
we obtain the formula
\begin{equation}
w_{eff} \left( a \right)\,
=\,-1\,-\frac{1}{3}\left(\frac{-4\Omega_{DR}\,+\,\nu\,\Omega_B\,a^{\nu+4}\,+\,\mu\,\Omega_5\,a^{\mu+4}\,}{
\Omega_{DR}\,+\,\Omega_B\,a^{\nu+4}\,+\,\Omega_5\,a^{\mu+4}\,}\right)\,,
\end{equation}
from which we can determine the variation of $w_{eff}$ with time.
As we see, all time dependence comes for the bulk terms. The
{\textit{deceleration parameter }} $q$ is given by
\begin{equation}
q \equiv  - \frac{1}{{H^2 }}\frac{{\ddot a}}{a} = \frac{{\Omega _m a^{ - 3}  + 2\Omega _{DR} a^{ - 4}  - \left( {\nu  + 2} \right)\Omega _B a^\nu   - \left( {\mu  + 2} \right)\Omega _5 a^\mu  }}{{2\left( {\Omega _m a^{ - 3}  + \Omega _{DR} a^{ - 4}  + \Omega _B a^\nu   + \Omega _5 a^\mu  } \right)}}\,.
\end{equation}
In the following treatment, we will re-express $q$ and $w_{eff}$
in terms of the redshift $z$, instead of the scale factor $a$. The
two are related by $a=\frac{a_0}{1+z}$, $a_0$ being the current
scale factor for our four-dimensional universe.

\section{Gold dataset fit}
We are going to use the updated Gold dataset (Gold06) compiled by
Riess et al. \cite{Riess:2006fw}, consisting of a total of 182
SnIa, located at distances ranging within the interval $0.024 \le
z \le 1.755$. This dataset allows a fairly accurate reconstruction
of the expansion history of our universe at late times.

The Gold dataset is compiled from various sources analyzed in a
consistent and robust manner with reduced calibration errors
arising from systematics. It contains 119 points from previously
published data plus 16 points with $0.46<z<1.39$ discovered
recently by the Hubble Space Telescope (HST). It also incorporates
47 points ($0.25<z<1$) from the first year release of the SNLS
dataset \cite{Astier:2005qq} out of a total of 73 distant SnIa.
Some supernovae were excluded \cite{Riess:2006fw} due to highly
uncertain color measurements or high extinction $A_V>0.5$, i.e.
the dimming of the SN magnitudes resulting from dust. A redshift
cut was also imposed at $c z< 7000km/s$ or $z<0.0233$, in order to
avoid the influence of a possible local ``Hubble Bubble", i.e. a
region with a higher Hubble parameter value due to the existence
of a large local void \cite{Jha:2006fm}. Thus, a high-confidence
subsample was defined.

The above observations provide the apparent magnitude $m(z)$ of
the supernovae at peak brightness after implementing the correction
for galactic extinction, the K-correction and the light curve
width-luminosity correction. The resulting apparent magnitude
$m(z)$ is related to the luminosity distance $d_L(z)$ through \be
m_{th}(z)={\bar M} (M,H_0) + 5 log_{10} (D_L (z))\,, \label{mdl} \ee
where in a flat cosmological model \be D_L (z)= (1+z) \int_0^z
dz'\frac{H_0}{H(z';a_1,...,a_n)} \label{dlth1} \,,\ee is the Hubble
free luminosity distance ($H_0 d_L/c$).
The parameters $a_1,...,a_n$ are
theoretical model parameters and ${\bar M}$ is the magnitude zero
point offset, depending on the absolute magnitude $M$ and on the
present Hubble parameter $H_0$ as \ba
{\bar M} &=& M + 5 log_{10}(\frac{c\; H_0^{-1}}{Mpc}) + 25= \nn \\
&=& M-5log_{10}h+42.38 \label{barm} \,.\ea The parameter $M$ is the
absolute magnitude and is assumed to be constant after the above
mentioned corrections have been implemented in $m(z)$.

The data points of the Gold06 dataset, after the corrections have
been implemented, are given in terms of the distance modulus as
\be \mu_{obs}(z_i)\equiv m_{obs}(z_i) - M \label{mug}\,.\ee The
theoretical model parameters are determined by minimizing the
quantity \be \chi^2 (a_1,...,a_n)= \sum_{i=1}^N
\frac{(\mu_{obs}(z_i) - \mu_{th}(z_i))^2}{\sigma_{\mu \; i}^2 +
\sigma_{v\; i}^2 } \label{chi2} \,,\ee where $\sigma_{\mu \; i}^2$
and $\sigma_{v\; i}^2$ are the errors due to flux uncertainties
and peculiar velocity dispersion respectively. These errors are
assumed to be gaussian and uncorrelated. The theoretical distance
modulus is defined as \be \mu_{th}(z_i)\equiv m_{th}(z_i) - M =5
log_{10} (D_L (z)) +\mu_0 \label{mth}\,, \ee where \be \mu_0=
42.38 - 5 log_{10}h \label{mu0}\,,\ee $\mu_{th}(z_i)$ depends also
on the parameters $a_1,...,a_n$ used in the parametrization of
$H(z)$ in equation (\ref{dlth1}).

To find the best fit parameters of the parametrization (\ref{H2}) we
used the prior $\omm=0.27$ and followed three approaches which we
label as {\it minimization}, {\it marginalization-minimization} and
{\it marginalization-averaging} with results which are consistent
among the three approaches. These approaches can be described as follows:

\begin{itemize}
    \item {\bf Minimization:} We minimized eq. (\ref{chi2}) with
    respect to all four parameters ($\Omega_{DR} $, $\tilde{\Omega}_B$,
    $\nu$ and $\mu$). The steps we followed for the minimization of
    (\ref{chi2}) are described in detail in Ref. \cite{compleg}.
    The advantage of this method is that it treats all parameters
    on an equal basis and gives direct information on all four of
    them. On the other hand it introduces significant degeneracy,
    thus leading to large error bars for the best fit parameter values.

    \item {\bf Marginalization-Minimization:} In this approach we
    marginalize over $\Omega_{DR} \in [-1.5, 5 ]$ and
    $\tilde{\Omega}_B \in [-10, 2]$ and minimize with respect to
    $\nu$ and $\mu$. To choose the range in which we will marginalize
    the two parameters, since we have no prior information
    for them, we minimized $\chi^2$ for several pairs of reasonable
    values for $\nu$ and $\mu$ and found a wide enough range that
    contains all the best fit $\Omega_{DR} $ and
    $\tilde{\Omega}_B$. Also, we have checked that our results are not affected
    if we choose a larger range than the one we used to
    marginalize. The marginalized $\tilde{\chi}^2$ is defined as:
    \be \tilde{\chi}^2(\nu,\mu)=-2~ln \int e^{-\chi^2/2} ~ d\Omega_{DR} ~
    d\tilde{\Omega}_B \label{margin} \,,\ee This is minimized with respect to
    $\nu$ and $\mu$. We thus obtain best fit values for the exponents
    $\nu $ and $\mu$ which are directly relevant to the physical
    properties of the two components $\bar{P}_B$ and $P_5$. This
    approach has less degeneracy at the minimum compared to the
    previous direct minimization but it does not provide an estimate
    for $\Omega_{DR} $ and $\tilde{\Omega}_B$ which are treated as
    nuisance parameters. Such an estimate is needed in order to
    construct the best fit effective equation of state parameter $w_{eff}(z)$,
    defined by eq. (\ref{eosp}). In order to obtain this estimate
    we fix $\nu$ and $\mu$ to the best fit values mentioned
    above and minimize $\chi^2$ with respect to $\Omega_{DR} $ and
    $\tilde{\Omega}_B$. \item {\bf Marginalization-Averaging:} The best
    fit values of $\Omega_{DR} $ and $\tilde{\Omega}_B$ obtained after
    marginalization as described above were also verified by
    considering the {\it average} values of these parameters instead
    of their values that minimize $\chi^2$. We thus define the average
    values of the two parameters as: \be < \Omega_{par} > = \frac{\int
    \Omega_{par} e^{-\chi^2/2} ~ d\Omega_{DR} ~
    d\tilde{\Omega}_B}{\int e^{-\chi^2/2} ~ d\Omega_{DR} ~
    d\tilde{\Omega}_B} \label{omegaver}\,, \ee where $par=\{\Omega_{DR} ,
    \tilde{\Omega}_B \}$.
\end{itemize}

\vspace{0pt}
\begin{table}[!b]
\begin{center}
\caption{The best fit parameter values obtained with each one of
the three methods. Note that the quantities that were determined
with the {\textit{Averaging}} procedure of eq. (\ref{omegaver}) were not
assigned an error due to the nature of the method. \label{table1}}
\begin{tabular}{cccccc}
\hline
\hline\\
\vspace{6pt}  \textbf{Method}     & $\nu$                       & $\mu $                         & $\Omega_{DR}$    & $\tilde{\Omega}_B$ &$\Omega_{5}$     \\
\hline \\
\vspace{6pt}  \textbf{Minimization}& $ 0.75\pm 5.09$\hspace{3pt}& $-3.55 \pm 5.77 $ \hspace{3pt} & $ 0.29 \pm 2.45$ & $3.39\pm 0.76$     & $-0.27\pm 2.57$ \\
\vspace{6pt}  \textbf{Marg.-Min.} & $ -3.0\pm 1.1$ \hspace{3pt} & $ -0.8 \pm 0.3 $ \hspace{3pt}  & $ 0.49 \pm 0.25$ & $-1.00 \pm 0.49$   & $1.24 \pm 1.23$ \\
\vspace{6pt}  \textbf{Marg.-Av.}  & $ -3.0\pm 1.1$ \hspace{3pt} & $ -0.8 \pm 0.3 $ \hspace{3pt}  & $ 0.52 $         & $-1.05$            & $1.26$          \\
\hline \hline
\end{tabular}
\end{center}
\end{table}

The best fit parameter values obtained with each one of the above
methods are shown in Table I. The first approach cannot be used to
draw a safe conclusion regarding the model under consideration due
to the large error-bars of the best-fit parameters. On the other
hand, the marginalization seems to hint towards $\nu \sim -3$ and
a negative $\tilde{\Omega}_B$, which corresponds to a positive
bulk pressure $\overline{P}_B$ (see eqs. (\ref{Ansatz}) and
(\ref{OmegaB})). Also, a positive $\Omega_5$ corresponds to
$P_5<0$ which means that energy is flowing from the bulk into the
brane. This is an intuitively reasonable result which may hint
towards a relativistic matter component which is propagating in
the bulk and is moving at a speed $v$ along the fifth dimension
\cite{BT,BDT}. The dark radiation term has been shown to be
related to the generalized comoving mass $\cal{M}$ of the bulk
fluid and its best-fit parameter $\Omega_{DR}$ is positive
\cite{Apostolopoulos:2004ic}. This a desirable feature of this
model, since a negative value would indicate a negative $\cal{M}$.
It also ensures a well-behaved cosmological evolution at early
times, although the validity of the effective Friedmann equation
in this regime is questionable. Finally, as we can see in Table I,
the two approaches, `Marginalization-Minimization' and
`Marginalization-Averaging' are in good agreement.

The corresponding forms of the equation of state parameter
$w_{eff}(z)$ and the deceleration parameter $q(z)$ are shown in
Figs 1 and 2 for the first two approaches (`Minimization' and
`Marginalization-Minimization'). Despite the large error-bars, the
best-fit forms of $w_{eff}(z)$ and $q(z)$ are in good agreement
between the two methods. The plots corresponding to the third
approach (`Marginalization-Averaging') are practically identical
with the Figs 1b and 2b of the `Marginalization-Minimization'
method (see also Table I).

\begin{figure*}[!t]
\begin{center}
\rotatebox{0}{\resizebox{1\textwidth}{!}{\includegraphics{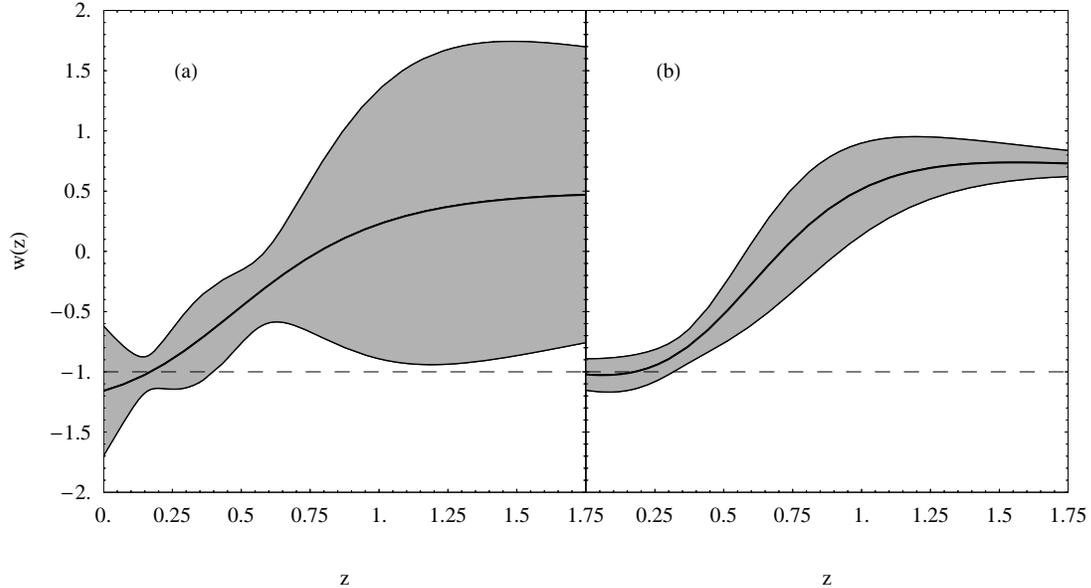}}}
\vspace{0pt}{ \caption{The effective equation of state $w_{eff}$
for the `Minimization' method (Fig. 1a) and the
`Marginalization-Minimization' approach (Fig. 1b). The large error
region of the in Fig. 1a is due to the significant degeneracy of
the model while the decreased error region in Fig. 1b is due to
the fact that the values of $\nu$ and $\mu$ were assumed fixed
during the minimization with respect to $\Omega_{DR} $ and
$\tilde{\Omega}_B$.}} \label{fig1}
\end{center}
\end{figure*}

A noteworthy feature of Figs 1a and 2a is the presence of
``sweet-spots'' at $z \sim 0.2$ and $z \sim 0.55$. The presence of
``sweet-spots'' at different redshifts for different
parameterizations, usually polynomial, is something which has been
studied previously (see \cite{Alam:2004ip}) and is a consequence
of the ansatz used. However, upon marginalization the error region
of Fig. 1b is significantly more smooth than the one of Fig. 1a.

\begin{figure*}[!t]
\begin{center}
\rotatebox{0}{\resizebox{1\textwidth}{!}{\includegraphics{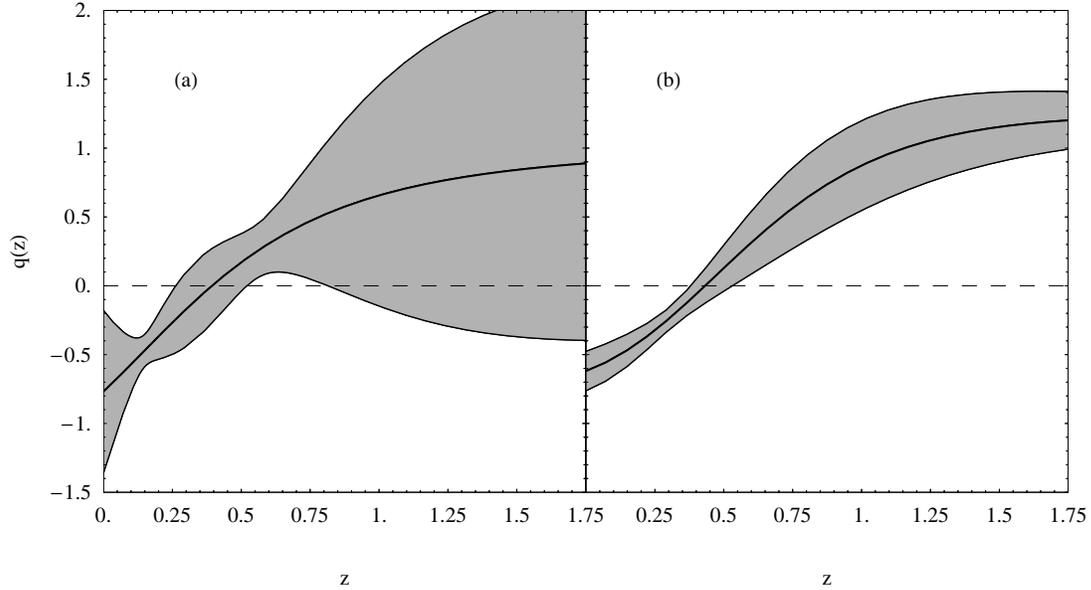}}}
\vspace{0pt}{ \caption{The deceleration parameter $q$ for the
`Minimization' method (Fig. 2a) and the
`Marginalization-Minimization' approach (Fig. 2b). The large error
region of the in Fig. 2a is due to the significant degeneracy of
the model while the decreased error region in Fig. 2b is due to
the fact that the values of $\nu$ and $\mu$ were assumed fixed
during the minimization with respect to $\Omega_{DR} $ and
$\tilde{\Omega}_B$.}} \label{fig2}
\end{center}
\end{figure*}

\section{Conclusion-Outlook}

We have used the latest SnIa data included in the Gold06 dataset
\cite{R} to provide an estimate of the preferred parameter values of
a braneworld model with bulk-brane energy exchange. The advantage of
this model is that it has a clear physical motivation based on
fundamental physics. Even though the large number of model
parameters that require fitting introduces large error bars, the
combination of different methods used for the fitting has provided
interesting hints for the preferred parameter values.

Due to the significant degeneracy of the model used, the first
fitting approach provided best fit parameter values with large
error-bars (see Table I) and does not allow for safe conclusions
to be drawn. On the other hand, the marginalization approach hints
towards a bulk energy component that behaves like relativistic
matter which is confined in the bulk and is moving at a speed $v$
along the fifth dimension while the bulk-brane energy exchange
component obtained corresponds to negative pressure, which means
that energy is flowing from the bulk into the brane. Despite of
these {\textit{natural values}} of parameters, the best fit \textit{effective}
equation of state parameter $w_{eff}$ marginally crosses the
phantom divide line $w=-1$. It also reproduces accelerated
expansion at late times, with a transition from acceleration to
deceleration around $z \approx 0.5$, consistent with previous
analysis \cite{Astier:2005qq,Riess:2006fw}. This demonstrates both
the ability of this class of braneworld models to provide crossing of the
phantom divide, while the cosmological data hint towards
{\textit{natural}} values for the model parameters.

{\bf Acknowledgements:} This work was supported by the European
Research and Training Network MRTPN-CT-2006 035863-1
(UniverseNet). S.N. acknowledges support from the Greek State
Scholarships Foundation (I.K.Y.). C. B. acknowledges also an
 {\textit{Onassis Foundation}} fellowship.

\end{document}